\documentclass[prl,showpacs,twocolumn,amsmath,nobibnotes]{revtex4-1}%revtex4-1}

\usepackage{cancel}
\usepackage[normalem]{ulem}

\usepackage{graphicx}
\usepackage{color}
\usepackage{array}
\usepackage{xcolor}
\usepackage[dvips]{epsfig}
\usepackage{epsfig}
\usepackage{enumerate}
\usepackage[hang, nooneline]{subfigure}
\usepackage{bbm, dsfont}
\usepackage{amsfonts,amsmath,amssymb,stmaryrd}

\newcommand{\bra}[1]{\langle #1 | \,}
\newcommand{\ket}[1]{\, | #1 \rangle}

\newcommand{\expv}[1]{\langle #1 \rangle}

\newcommand{\bl}{\begin{linenomath*}}
\newcommand{\el}{\end{linenomath*}}
\newcommand{\E}{\hat{\mathcal{E}}}
\newcommand{\s}[1]{\hat\sigma_{#1}}
\newcommand{\pdt}{\frac{\partial}{\partial t}}
\newcommand{\pdz}{\frac{\partial}{\partial z}}

\newcommand{\be}{\begin{equation}}
\newcommand{\ee}{\end{equation}}
\newcommand{\bea}{\begin{eqnarray}}
\newcommand{\eea}{\end{eqnarray}}
\renewcommand{\vec}[1]{\mathbf{#1}}

\newcommand{\D}{\hat\Psi}

\newcommand{\e}{\text{e}}

\begin{document}

% \setpagewiselinenumbers
%\modulolinenumbers[5]
%\linenumbers

%%%%%%%%%%%%%%%%%%%%%%%%%%%%%%%%%%%%%%%%%%%%%%%%%%
\title{Spinor Slow-Light and Dirac particles with variable mass}
%%%%%%%%%%%%%%%%%%%%%%%%%%%%%%%%%%%%%%%%%%%%%%%%%%

%\author{R.G. Unanyan$^{1}$,  J. Otterbach$^{1}$,  M. Fleischhauer$^{1}$}
\author{R.G. Unanyan$^{1}$}
\author{J. Otterbach$^{1}$}\email[corresponding author: ]{jotterbach@physik.uni-kl.de}
\author{M. Fleischhauer$^{1}$}
\affiliation{$^{1}$Department of Physics and research center OPTIMAS, Technische Universit\"at Kaiserslautern, 67663 Kaiserslautern, Germany}
\author{J. Ruseckas$^{2}$, V. Kudria\v{s}ov$^{2}$ G. Juzeli\=unas$^{2}$}
\affiliation{$^{2}$Institute of Theoretical Physics and Astronomy, Vilnius University, A. Go\v{s}tauto 12, 01108 Vilnius, Lithuania}

\date{\today}

%%%%%%%%%%%%%%%%%%%%%%%%%%%%%%%%%%%%%%%%%%%%%%%%%%
\begin{abstract}
We consider the interaction of two weak probe fields of light with an atomic ensemble coherently driven by two pairs of standing wave laser fields in a tripod-type linkage scheme. The system is shown to exhibit a Dirac-like spectrum for light-matter quasi-particles with multiple dark-states, termed spinor slow-light polaritons (SSP). They posses an ``effective speed of light`` given by the group-velocity of slow-light, and can be made massive by inducing a small two-photon detuning. Control of the two-photon detuning can be used to locally vary the mass including a sign flip.
This allows e.g. the implementation of the random-mass Dirac model for which localized zero-energy (mid-gap) states exist with unsual long-range correlations.
\end{abstract}
%%%%%%%%%%%%%%%%%%%%%%%%%%%%%%%%%%%%%%%%%%%%%%%%%%%

\pacs{42.50.Ct, 03.65.Pm, 71.23.An}

\maketitle
%%%%%%%%%%%%%%%%%%%%%%%%%%%%%%%%%%%%%%%%%%%%%%%%%%%
%%%%%%%%%%%%%%%%%%%%%%%%%%%%%%%%%%%%%%%%%%%%%%%%%%%

%%%%%%%%%%%%%%%%%%%%%%%%%%%%%%%%%%%%%%%%%%%%%%%%%%%
%%%%%%%%%%%%%%%%%%%%%%%%%%%%%%%%%%%%%%%%%%%%%%%%%%%
% \section{introduction}
%%%%%%%%%%%%%%%%%%%%%%%%%%%%%%%%%%%%%%%%%%%%%%%%%%%
%%%%%%%%%%%%%%%%%%%%%%%%%%%%%%%%%%%%%%%%%%%%%%%%%%%

Due to the recent experimental and theoretical progress on various system there has been a growing interest in
interacting systems with effective single-particle Dirac dynamics. A prime example is Graphene \cite{Novoselov-RMP-2009} which shows unusual low-energy properties, described by Dirac quasi-particles. The effective ``speed of light'' of these quasi-particles is  given by the electron Fermi velocity, which is about $300$ times smaller than the vacuum speed of light.
We here show that an effective Dirac-like dynamics also emerges for two-component, i.e. spinor-like,  slow-light polaritons \cite{Fleischhauer-RMP-2005} in 1D.  They possess a controllable effective speed, corresponding to the much
smaller group velocity of light \cite{Hau-Nature-1999} in media exhibiting electromagnetically induced transparency (EIT)\cite{Fleischhauer-RMP-2005}. This may allow experimental studies of a number of interesting effects of relativistic dynamics at low energies and small velocities. Additionally the effective mass, determined by laser detuning, is locally and dynamically adjustable including the sign, which provides access to interesting phenomena such as the unusual localization in a Dirac model where the mass is a random function of space \cite{Balents-PRB-1997,Tsvelik-PRB-1998}. The latter is  inaccessible for truly massive particles such as cold atoms.

\begin{figure}[t]
\includegraphics[width=.4\textwidth]{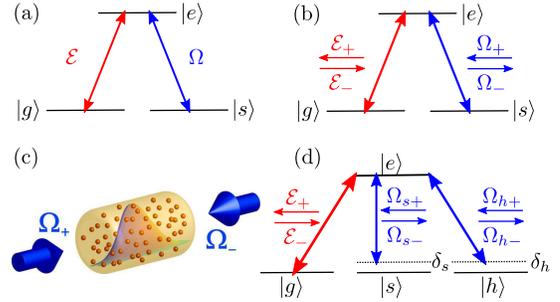}
 \caption{(Color online) (a) 3-level $\Lambda$-type linkage pattern for the creation of slow-light polaritons. The strong control field $\Omega$ produces electromagnetically induced transparency (EIT) for the weak probe field $\mathcal E$. (b) Linkage pattern with two counter-propagating control fields $\Omega_\pm$ to create a stationary pattern of counter-propagating probe light ${\mathcal E}_\pm$. (c) Typical experimental setup for the creation of single-component stationary light by use of two counter-propagating, mutually orthogonal control fields $\Omega_\pm$. (d) Tripod-type linkage pattern for the creation of spinor slow-light polaritons (SSP).}
 \label{fig:LinkagePatterns}
\end{figure}

Slow-light polaritons are formed in the Raman interaction of a weak probe field with a coherently driven ensemble of atoms with a  $\Lambda$-type linkage pattern (Fig. \ref{fig:LinkagePatterns}a). They build the basis of ultra slow light  \cite{Hau-Nature-1999}, light storage \cite{Phillips-2001,Liu-01} and, upon using two counter-propagating control fields (Fig.\ref{fig:LinkagePatterns}b,c), stationary light \cite{Bajcsy-nature-2003,Zimmer-OC-2006,Lin-PRL-2009}. They exhibit extraordinary long lifetimes ranging from several hundreds of microseconds\cite{Phillips-2001} up to hundreds of millisecond \cite{Schnorrberger-PRL-2009,Hau-PRL-2009} and offer a wide range of tuning parameters.

%%%%%%%%%%%%%%%%%%%%%%%%%%%%%%%%%%%%%%%%%%%%%%%%%%%%
%%%%%%%%%%%%%%%%%%%%%%%%%%%%%%%%%%%%%%%%%%%%%%%%%%%%

In the following we argue that by adding additional states to the above linkage pattern (Fig.\ref{fig:LinkagePatterns}d), we can create a \textit{spinor-like} object consisting of \textit{two} adiabatic eigensolutions immune to spontaneous decay. The suggested tripod linkage interaction between atoms and lights is a minimal realization of spinor slow-light polaritons (SSP). They obey an effective 1D Dirac equation, thus creating relativistic quasi-particles with an internal degree of freedom. They travel at an effective speed $c^*$ given by the slow-light group velocity, and posses a mass $m^*$ that is determined by a variable two-photon detuning. The possibility of a locally adjustable mass, which is absent for any truly massive particle, allows to study a number of interesting phenomena:  For example, if the mass of the Dirac particle is a randomly varying function of space with a vanishing mean-value,  there exists a mid-gap (zero-energy) state with unusual correlations. Random-mass Dirac Hamiltonians describe a number of effects in condensed-matter systems, ranging from disordered half-filled metals \cite{Gogolin-PhysRep-1982},  random anti-ferromagnetic spin-1/2 chains \cite{Fisher-PRB-1994,McKenzie-PRL-1996,Balents-PRB-1997}, random transverse-Ising spin-1/2 chains \cite{Fisher-PRB-1995,Fisher-PRL-1998}, spin Peierls and spin-ladder systems  \cite{Fabrizio-PRL-1997,Tsvelik-PRB-1998,Steiner-PRB-1998}.  Many aspects of random-mass Dirac systems are not understood, e.g. the effect of local impurities or boundaries \cite{Texier-JPA-2010} or when interactions are added. Spinor-polaritons may offer new experimental access to these issues.

%%%%%%%%%%%%%%%%%%%%%%%%%%%%%%%%%%%%%%%%%%%%%%%%%%%%
\begin{figure}
\centering
\includegraphics[width=.43\textwidth]{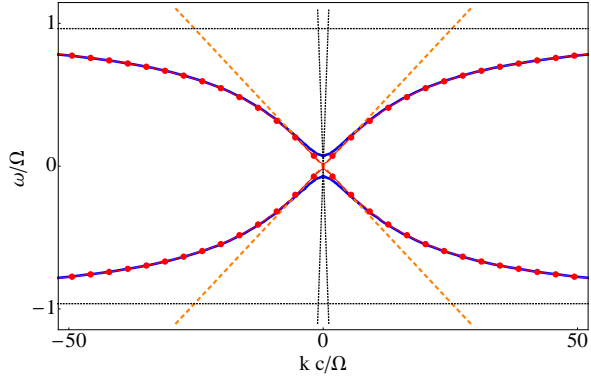}
 \caption{(Color online)
Dispersion relation of the double-tripod pattern (Fig. \ref{fig:LinkagePatterns}d) (ignoring decay) showing the energy branches of the spinor slow-light polaritons (SSP) (red dotted line: $\delta=0$, blue solid line: $\delta=0.075\Omega_0$ ).}
 \label{fig:dispersion_relation}
\end{figure}

%%%%%%%%%%%%%%%%%%%%%%%%%%%%%%%%%%%%%%%%%%%%%%%%%%%%%%%%%%%%%
We here consider light coupled to an ensemble of atoms with a linkage pattern as seen in Fig. \ref{fig:LinkagePatterns}d, consisting of three (meta-)stable states $\ket g$, $\ket s$, $\ket h$ and one excited state $\ket e$ with a decay rate $\gamma$. There are two pairs of counter-propagating control fields with Rabi-frequencies $\Omega_{s\pm}\e^{\pm ik_sz}$ and $\Omega_{h\pm}\e^{\pm ik_hz}$ on the transitions $\ket{s}-\ket{e}$ and $\ket{h}-\ket{e}$, respectively, as indicated in Fig. \ref{fig:LinkagePatterns}d. These create EIT  for a pair of counter propagating probe fields  $\hat E_\pm$, coupling the transition $\ket{g}-\ket{e}$ \cite{Fleischhauer-RMP-2005}. The use of pairs of counter-propagating control beams is essential to produce two-component slow- or stationary light, a feature missing in the ordinary tripod-type linkage pattern \cite{Paspalakis-2002, Zaremba07PRA,RusecMekJuz10}. This scheme effectively creates two parallel tripod-type linkage patterns sharing common ground-states. We introduce amplitudes $\E_\pm(z,t)$ of the probe fields that vary slowly in space and time by $\hat E_\pm=\sqrt{\frac{\hbar\omega_p}{2\varepsilon_0 V}}\E_\pm(z,t)\exp\{-i(\omega_pt\mp k_pz)\}+h.a.$. Furthermore we define continuous flip operators via $\s{\mu\nu}(z,t)=\frac{1}{\Delta N}\sum_{j\in \Delta V(z)}\s{\mu\nu}^j$, with $\s{\mu\nu}^j\equiv\ket{\mu}_{jj}\bra{\nu}$ being the flip operator for the $j$-th atom and the sum being taken over $\Delta N$ atoms in a volume $\Delta V(z)$.

The full dynamics of the system is governed by the Maxwell-Bloch equations (MBE) in 1D. In the weak field limit the ground-state is only slightly depleted allowing to assume $\s{gg}\approx 1$ and $\s{\mu\nu}\approx 0$ in lowest order of perturbation. We further expand the relevant coherence $\sigma_{ge}=\sigma_{ge}^{(+)}\e^{ikz}+\sigma_{ge}^{(-)}\e^{-ikz}$, and neglect all higher spatial $k$-components as well as couplings to other coherences with higher Fourier $k$-modes. This approximation is know as the secular approximation \cite{Zimmer-OC-2006,Moiseev-2006}, and is well justified in hot atomic gases \cite{Fleischhauer-1994,Bajcsy-nature-2003}. We can thus write down the linearized Maxwell-Bloch equation (MBE)  in form of a $6\times 6$ matrix consisting of $2\times 2$ sub-blocks
\begin{align}
 \pdt\begin{pmatrix}
  \pmb\E \\ \hat{\pmb\Sigma} \\ \hat{\vec P}
 \end{pmatrix}
\,=\,\begin{pmatrix}
      -c\underline K  & 0  & i\underline G \\
      0  & i\underline\delta  &i\underline\Omega^\dagger \\
      i\underline G & i\underline\Omega & -\underline\Gamma  \\
     \end{pmatrix}
\begin{pmatrix}
  \pmb\E \\ \hat{\pmb\Sigma} \\ \vec{\hat P}
 \end{pmatrix}+\hat{\vec{F}}_P,
\label{eq:HamiltonianPhysicalBasis}
\end{align}
where we defined a two component field vector $\pmb\E=(\E_+,\E_-)^T$, a two-component vector of spin polarizations  $\hat{\pmb\Sigma}=(\sqrt{n}\s{gs},\sqrt{n}\s{gh})^T$, and a two-component vector of optical polarizations $\vec{\hat P}=(\sqrt{n}\s{ge_+}, \sqrt{n}\s{ge_-})^T$. Here $\underline K=\sigma_{z}\pdz$ and $\underline G=g\sqrt{n}\;\mathds{1}_{2\times 2}$, where $g=\wp\sqrt{\omega/2\hbar\varepsilon_{0}}$ is the coupling constant of $\E_\pm$ to the transition $\ket{g}-\ket{e}$ with dipole matrix element $\wp$ and $n$ is the atomic number density. Furthermore $\underline\delta=\text{diag}\{\delta_s,\delta_h \}$, $\underline\Omega=\begin{pmatrix}\Omega_{s+} & \Omega_{h+} \\   \Omega_{s-} & \Omega_{h-} \\ \end{pmatrix}$, $\underline\Gamma=\text{diag}\{\gamma_{ge_+}-i\Delta_+,\gamma_{ge_-}-i\Delta_- \}$. Finally $\hat F_P$ are Langevin noise forces neccessary for preserving the commutation relations of the decaying variables $\hat{\mathbf{P}}$. For an exponential decay these operators are $\delta$-correlated in time, i.e. $\expv{\hat F_A(t)\hat F_B(t')}=D_{AB}\delta(t-t')$, where the diffusion coefficients $D_{AB}$ are proportional to the population of the excited states and can be calculated by means of the dissipation-fluctuation theorem \cite{Louisell-1973}. With the help of a generalized Morris-Shore transformation \cite{Zimmer-PRA-2008}, one can show that in linear response this system has {\em two} adiabatic eigensolution that are decoupled from the excited states. If we choose the control fields such that  $\underline \Omega= \frac{\Omega_0}{2}(\mathds{1}+i\sigma_x)$, where $\sigma_x$ is a Pauli matrix, the eigensolution of the tripod systems, the dark-state polaritons   (DSP), are
\begin{align}
\hat\Psi_{+} &  =\cos\theta \E_{+}-\frac{1}{\sqrt{2}}\sin\theta\left(
\hat\sigma_{gs}-i\hat\sigma_{gh}\right), \\
\hat\Psi_{-} &  =\cos\theta \E_{-}+\frac{1}{\sqrt{2}}\sin\theta\left(
i\hat\sigma_{gs}-\hat\sigma_{gh}\right),
\label{eq:darkstates}
\end{align}
where $\tan^2\theta=g^2n/\Omega_0^2$ is the mixing angle between light and matter excitation. We emphasize that in contrast to ordinary stationary light where only {\it one} independent dark polariton  exists \cite{Zimmer-PRA-2008} we here have two independent dark eigensolutions.
We now introduce a small two-photon detuning of opposite sign   $\delta=\delta_s=-\delta_h$  for the transitions $\ket{g}-\ket{s}$ and $\ket{g}-\ket{h}$, respectively, with $\left|\delta\right|\ll \Omega_0^2/\gamma$. Assuming one-photon resonance, i.e. $\Delta_\pm=0$, setting $\gamma_{ge_\pm}=\gamma$, and transforming eq. (\ref{eq:HamiltonianPhysicalBasis}) to the polariton basis and adiabatically eliminating all variables except for the DSPs, we arrive at
\begin{align}
i\hbar\frac{\partial}{\partial t}\mathbf{\D}\,&=\,\left(i\hbar v_\text{g}\sigma_{z}\frac{\partial}{\partial }+\hbar\delta\sin^{2}\theta\sigma_{y}\right)\mathbf{\D}\nonumber \\
 &\,-iL_\text{abs}v_\text{g}\sin^4\theta\left(\sigma_z\frac{\partial}{\partial z}-\frac{\delta}{c}\sigma_y\right)^2\mathbf{\D}+\hat F_\Psi.
\label{eq:SSP}
\end{align}
Eq. (\ref{eq:SSP}) represents a Dirac equation for a spinor $\mathbf{\D}=\bigl(\D_+,\D_-\bigr)^T$ with effective ''speed of light'' and mass
\begin{align}
 c^* = v_{\rm g}=c\cos^2\theta,\qquad m^*=\frac{\hbar\delta\sin^2\theta}{v_\text{g}^2}.
\end{align}
The imaginary quadratic term in (\ref{eq:SSP}) results from non-adiabatic couplings to decaying states, $\hat F_\Psi$ is the corresponding Langevin-noise operator. $L_\text{abs}=c\gamma/g^2n$ is the resonant absorption length in absence of EIT,  $\sigma_z$ and $\sigma_y$ denote Pauli matrices. From eq. (\ref{eq:SSP}) we expect a Dirac cone structure of the dispersion relation as seen in Fig.\ref{fig:dispersion_relation}. The interaction of the control fields with the atoms results in two Autler-Townes states represented by the horizontal black dashed lines. The dispersion of the free probe fields is given by the almost vertical black dashed lines. Switching on the interaction, the red dotted dispersion branches are formed. These correspond to the two components of the slow-light spinor and are in a very good approximation described by the expected Dirac dispersion with $c^*=v_\text{g}$ around $k=0$ (orange dashed line). Finally inducing a small two-photon detuning $\delta$ results in a finite mass $m^*$, as can be seen by the finite splitting of the blue (solid) line.

To confirm the effective model, we numerically integrated the full set of MBE  for the scheme of Fig. \ref{fig:LinkagePatterns}d in time for the case of  $m^*=0$. The results showed very good agreement with the analytic prediction.

\begin{figure}[t!]
 \centering
  \includegraphics[width=1\columnwidth]{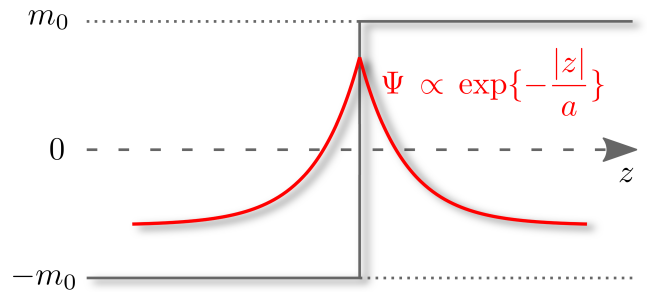}
 \caption{(Color online) Zero energy bound state at a mass jump $m(z) = m\, {\rm sgn}(z)$. $a=\lambda_c= \hbar/(m^* c^*)$ is the
effective Compton length.}
\label{fig:localized-state}
\end{figure}

We now exploit the property of a locally adjustable mass of the SSP. If at a certain point along the $z$-axis the mass changes its sign, i.e. $ m_0\rightarrow -m_0$, one can easily show, that there exists a localized mid-gap (i.e. zero energy) solution around the mass jump, as by squaring the Dirac Hamiltonian leads to a pair of supersymmetric Hamiltonians corresponding to two independent particles, described by a Schr\"odinger equations with a Dirac-$\delta$ potential. The zero-energy bound state is exponentially localized with a localization length proportional to the effective Compton length $\lambda_C^* = \hbar/(m^* c^*)$. This is indicated in Fig. \ref{fig:localized-state}.
%%%%%%%%%%%%%%%%%%%%%%%%%%%%%%%%%%%%%%%%%%%%%%%%%%%%%%%%%%%%%
%%%%%%%%%%%%%%%%%%%%%%%%%%%%%%%%%%%%%%%%%%%%%%%%%%%%%%%%%%%%%

Thus one can ask how the state will look like when the number of mass jumps and their size is increased? This leads to the 1D random-mass Dirac model: Given a spatially random mass $m(z)$, with ($\hbar=c=1$)
\begin{align}
 \overline{m(z)m(y)}\,=\,2\Gamma \delta(z-y),\quad \overline{m(z)}\,=\,0,
 \label{eq:WhiteNoiseMass}
\end{align}
where $\Gamma$ parameterizes the strength of the disorder, it was shown \cite{Balents-PRB-1997,Tsvelik-PRB-1998} that the density of states at zero energy is diverging, corresponding to a localized state. The corresponding wave function $\Psi(z)$ exposes however unusual density correlations. Contrary to expectations there is not an exponential localization as in the case of Fig.\ref{fig:localized-state} but one finds for $\Gamma z\gg 1$:
\begin{align}
 \overline{\left|I(z)I(0)\right|}\,\sim \, \left(\frac{1}{\Gamma z} \right)^{3/2},
\label{eq:intensity-intensity-correlation}
\end{align}
where $I(z)=|\mathbf{\Psi}(z)|^2$.  A spatially varying mass (detuning $\delta$) can be achieved by using fluctuating magnetic fields or by applying speckle patterns  as done for the demonstration of Anderson localization \cite{Billy-Nature-2008}.

In experiments one has a finite correlation length $\xi$ of the disorder which  we use as a discretization length  for the simulations. We assume that the two-photon detuning has a Gaussian distribution with  $\overline{\delta}=0$ and $\overline{\delta^2}=\sigma^2$ for all points in space.

%%%%%%%%%%%%%%%%%%%%%%%%%%%%%%%%%%%%%%%%%%%%%%%%%%%%%%%%%%%%%
\begin{figure}[t!]
 \centering
  \includegraphics[width=.5\textwidth]{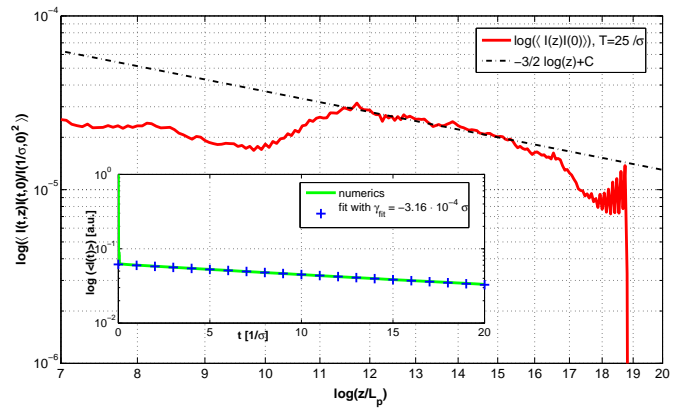}
 \caption{(Color online) Correlation function $\overline{\left|I(z)I(0)\right|^2}$ with disorder width $\sigma=0.01\gamma$ averaged over 50 realizations. or large times, the numerical result show very good agreement with the theoretical expected power-law (black dashed line), with the only fit-parameter being the ordinate intersection. The inlet shows the  intensity $\overline{I(t)}$ spatially integrated over a region $\pm 3 L_p$. The curve is normalized to the initial intensity $I(t_0,z=0)$. The fitted curve (blue crosses) corresponds to the theoretically predicted  effective decay rate with $L_*\approx\sqrt{2}L_p$.}
\label{fig:intensity-intensity-correlation}
\end{figure}
%%%%%%%%%%%%%%%%%%%%%%%%%%%%%%%%%%%%%%%%%%%%%%%%%%%%%%%%%%%%%
In Fig. \ref{fig:intensity-intensity-correlation} numerical results of the density correlation (\ref{eq:intensity-intensity-correlation}) are shown
obtained from the full MBE.  For large times the correlation (red solid line) approaches the predicted power-law behavior, shown by the black dashed line.

The finite two-photon detuning also leads to some small losses, seen in the inlet of Fig. \ref{fig:intensity-intensity-correlation}. Two regimes are apparent: The first rapid loss is due to imperfect matching of the initial wave function to the zero-energy eigenstate.  The smaller losses for large times can be attributed to non-adiabatic couplings of the SSP to decaying states, resulting in an effective loss rate $\gamma_\text{eff}=\pi^2L_\text{abs}v_\text{g}/L_*^2$, $L_*$ being some characteristic length scale of the localized state. These losses are negligible as long as $\gamma_\text{eff}T\ll 1$, where  $T=L/v_\text{g}$ is the typical time scale of the experiment. Furthermore (\ref{eq:intensity-intensity-correlation}) requires $\Gamma L\gg 1$, in order to observe the power-law decay. This translates to $L\gg v_\text{g}^2 n_\text{kinks}/\sigma^2\sin^4\theta\equiv L_\text{corr}$, where $n_\text{kinks}$ is the number density of mass jumps. Both equations finally lead to $L/L_{\text{abs}}\equiv OD>>\max \left\{ \frac{\Omega _{0}^{2}}{\gamma
\sigma }\sqrt{N_{\text{kinks}}},\left( \pi L/L_{\ast }\right) ^{2}\right\} $
. Here $N_\text{kinks}=n_\text{kinks}L$ is the number of kinks  in the length $L$ and $OD$ denotes the optical depth of the medium. For white noise the mean free path $\lambda_\text{mean}=n_\text{kinks}^{-1}$ can be approximated by the disorder correlation length $\xi$, i.e. $\lambda_\text{mean}\approx\xi$. Taking $\xi\approx0.3\,\mu$m,  $L\approx 4$ mm \cite{Billy-Nature-2008}, and estimating  $N_\text{kinks}=L/\xi\approx 10^4$ yields a required optical depth of $OD\geq 100$. This high optical depth is however achievable in systems such as atoms trapped in a hollow core photonic bandgap fiber \cite{Bajcsy-PRL-2009} or cold atoms trapped in the evanescent fields of an ultra-thin optical fiber \cite{Vetsch-2009}.

An estimate of the minimal build-up time of the correlations, defined as $T_\text{corr}=L_\text{corr}/v_\text{g}$ yields $T_\text{corr}\approx 2.5-25 \mu$s. Here we used $\sigma\approx0.1\gamma=(2\pi)\,0.6$MHz (for $^{87}$Rb), $v_\text{g}\approx 10-100$m/s \cite{Hau-Nature-1999}  and the correlation length $\xi\approx 0.3\mu$m from above. This time is far shorter than the achievable lifetime of the dark-state polaritons, typically on the order of 100$\mu$s \cite{Phillips-2001} to several ms \cite{Schnorrberger-PRL-2009,Hau-PRL-2009}, thus enabling experimental observation.

A possible experimental protocol to observe the proposed effect is to store an initial wave packet in the atomic coherences with existing storing techniques for slow-light polaritons \cite{Phillips-2001} or by coherent RF-transitions \cite{Li-2009}. Subsequently one applies the disorder, e.g. by a laser-speckle induced ac-Stark shifts and reads out the polaritons by applying all four control fields. The polariton states then evolve according to the above presented model. After a certain time the disorder is removed and the final state is read out by means of usual slow-light. Looking at the time profile of the intensity of the probe fields arriving at the detector one can reconstruct the spatial intensity profile of the localized state.

In summary we have shown that the coupling of two counter-propagating light fields to an optical thick ensemble of atoms driven by multiple drive fields in a tripod-like linkage pattern leads to the formation of spinor-like eigensolutions which are immune to spontaneous emission losses. These spinor dark-state polaritons obey a 1D Dirac equation with effective speed  $c^*$ given by the group velocity of slow-light in EIT media and with an effective mass $m^*$ determined by a small two-photon detuning. This allows to experimentally study a number of interesting phenomena of relativistic quantum dynamics. For example the possibility to tune the effective mass can be used to observe the unusual localization transition predicted for the random-mass Dirac model, which appears in the context of several interesting condensed matter systems with disorder. Furthermore adding interactions, e.g. by means of resonantly enhanced Kerr nonlinearities \cite{Lukin-Nature}, important relativistic many-body models such as the bosonic counterpart of the Thirring model \cite{Thirring-1958} should be experimentally accessible.

%%%%%%%%%%%%%%%%%%%%%%%%%%%%%%%%%%%%%%%%%%%%%%%%%%%%%
\
\textit{Acknowledgments. --} This work was supported by the DFG through the GRK 792 and the project UN 280/1, as well as by the Alexander von Humboldt Foundation, the Research Council of Lithuania and the EU project STREP NAMEQUAM.
%%%%%%%%%%%%%%%%%%%%%%%%%%%%%%%%%%%%%%%%%%%%%%%%%%%%%

%%%%%%%%%%%%%%%%%%%%%%%%%%%%%%%%%%%%%%%%%%%%%%%%%%%%%%

\begin{thebibliography}{99}


\bibitem {Novoselov-RMP-2009}A. H. Castro Neto, F. Guinea, N. M. R. Peres, K. S. Novoselov, and A. K. Geim, Rev. Mod. Phys. \textbf{81}, 109--162 (2009).

%\bibitem{Kempe-2003}J. Kempe, Contemp. Phys.  \textbf{44}, 307-327 (2003).

%\bibitem{Bracken-2007}A. J. Bracken, D. Ellinas, and I. Smyrnakis, Phys. Rev. A \textbf{75}, 022322 (2007).

\bibitem{Fleischhauer-RMP-2005}M. Fleischhauer, A. Imamoglu, and J. P. Marangos, Rev. Mod. Phys. \textbf{77}, 633--673 (2005).

\bibitem{Hau-Nature-1999}L. V. Hau, S. E. Harris, Z. Dutton, and C. H. Behroozi, Nature \textbf{397}, 594--598 (1999).

\bibitem{Balents-PRB-1997}L. Balents, and M. P. A. Fisher, Phys. Rev. B, \textbf{56}, 12970--12991 (1997).

\bibitem{Tsvelik-PRB-1998}D. G. Shelton, and A. M. Tsvelik, Phys. Rev. B, \textbf{57}, 14242--14246 (1998).

% \bibitem{Shelton-PRB-1996}D. G. Shelton and A. M. Tsvelik, Phys. Rev. B, \textbf{53}, 14036-14039 (1996).

\bibitem{Phillips-2001}D. F. Phillips, A. Fleischhauer, A. Mair, R. L. Walsworth, and M. D. Lukin, Phys. Rev. Lett. \textbf{86}, 783-786 (2001).

\bibitem{Liu-01}C.~Liu \emph{et al.}, Nature \textbf{409}, 490 (2001).

\bibitem{Zimmer-OC-2006}F. Zimmer, A. Andre, M. D. Lukin, and M. Fleischhauer, Opt. Comm. \textbf{264}, 441-453 (2006).

\bibitem{Bajcsy-nature-2003}M. Bajcsy, A. S. Zibrov, and M. D. Lukin, Nature \textbf{426}, 638-641 (2003).

\bibitem{Lin-PRL-2009} Yen-Wei Lin {\it et al.}, Phys. Rev. Lett. \textbf{102}, 213601 (2009).

\bibitem{Schnorrberger-PRL-2009}U. Schnorrberger, \textit{et. al}, Phys. Rev. Lett. \textbf{103}, 033003 (2009).

\bibitem{Hau-PRL-2009}  R. Zhang, S.R. Garner, and L. V. Hau, Phys. Rev. Lett. \textbf{103}, 233602 (2009)

\bibitem{Gogolin-PhysRep-1982} A. A. Gogolin, Phys. Rep. \textbf{86}, 1-53 (1982).

\bibitem{Fisher-PRB-1994} D. S. Fisher, Phys. Rev. B \textbf{50}, 3799 (1994)

\bibitem{McKenzie-PRL-1996} R. H. McKenzie, Phys. Rev. Lett. \textbf{77},4804 (1996).

\bibitem{Fisher-PRB-1995} D. S. Fisher, Phys. Rev. B \textbf{51}, 6411 (1995).

\bibitem{Fisher-PRL-1998} D.S. Fisher {\it et al.}, Phys. Rev. Lett. \textbf{80}, 3539 (1998).

\bibitem{Fabrizio-PRL-1997}M. Fabrizio,and R. M$\acute{\text{e}}$lin, Phys. Rev. Lett., \textbf{78}, 3382--3385 (1997).

\bibitem{Steiner-PRB-1998} M. Steiner, M. Fabrizio, and A.O.  Gogolin, Phys. Rev. B \textbf{57}, 8290 (1998).

\bibitem{Texier-JPA-2010} C. Texier, and C. Hagendorf, J. Phys. A \textbf{43}, 025002 (2010).

\bibitem{Paspalakis-2002}E. Paspalakis, N. J. Kylstra, and P. L. Knight, Phys. Rev. A \textbf{65}, 053808 (2002).

\bibitem{Zaremba07PRA}A. Raczynski , J. Zaremba , S. Zielinska-Kaniasty, Phys. Rev. A \textbf{75}, 013810 (2007).

\bibitem{RusecMekJuz10}J. Ruseckas, A. Mekys, and G. Juzeli\=unas, Optics and Spectroscopy \textbf{108}, 438 (2010).

\bibitem{Moiseev-2006}S. A. Moiseev, and B. S. Ham, Phys. Rev. A \textbf{73}, 033812 (2006).

\bibitem{Fleischhauer-1994}M. Fleischhauer, M. D. Lukin, D. E. Nikonov, and M. O. Scully, Opt. Comm. \textbf{110}, 351-357 (1994).

%\bibitem{SOM}See EPAPS Document No. \textit{[number will be inserted by publisher]} for a detailed descriptions and calculations.

% \bibitem{Fleischhauer-PRL-2008}Fleischhauer, M., Otterbach, J. \& Unanyan, R. G. Bose-Einstein Condensation of Stationary-Light Polaritons. Phys. Rev. Lett. \textbf{101}, 163601 (2008).

% \bibitem{Otterbach-PRL-2009}Otterbach, J., Unanyan, R. G. \& Fleischhauer, M. Confining Stationary Light: Dirac Dynamics and Klein Tunneling. Phys. Rev. Lett. \textbf{102}, 063602 (2009).

% \bibitem{Otterbach-2009}Otterbach, J., Ruseckas, J., Unanyan, R. G., Juzeli\=unas, G. \& Fleischhauer, M. Effective magnetic fields for stationary light. Phys. Rev. Lett. \textbf{104}, 033903 (20109.

\bibitem{Louisell-1973}W. H. Louisell, \textit{Quantum Statistical Properties of Radiation} Ch.7 (John Wiley \& Sons, New York, 1973).

\bibitem{Zimmer-PRA-2008}F. E. Zimmer, J. Otterbach, R. G. Unanyan, B. W. Shore, and M. Fleischhauer, Phys. Rev. A \textbf{77}, 063823 (2008).

% \bibitem {Calogeracos-CP-1999}A. Calogeracos, and N. Dombey, Contemp. Phys. \textbf{40}, 313 (1999).

% \bibitem {Unanyan-OC-1998}Unanyan, R., Fleischhauer, M., Shore, B. W. \& Bergmann, K. Robust creation and phase-sensitive probing of superposition states via stimulated Raman adiabatic passage (STIRAP) with degenerate dark states. Opt. Comm. \textbf{155}, 144--154 (1998).

\bibitem{Billy-Nature-2008}J. Billy \textit{et al.}, Nature \textbf{453}, 891--894 (2008).

\bibitem{Bajcsy-PRL-2009}M. Bajcsy \textit{et al.}, Phys. Rev. Lett. \textbf{102}, 203902 (2009).

\bibitem{Vetsch-2009}E. Vetsch \textit{et al.}, Preprint available at (http://arxiv.org/abs/0912.1179) (2009).

\bibitem{Li-2009}H. B. Li \textit{et al.}, Phys. Rev. A \textbf{80}, 023820 (2009).

\bibitem{Lukin-Nature} M.D. Lukin and A. Imamoglu, Nature \textbf{413} 6853 (2001).

\bibitem{Thirring-1958}W. Thirring, Ann. Phys. \textbf{3}, 91 (1958).

% \bibitem{Zaehringer-2010}F. Z\"ahringer \textit{et al.}, Phys. Rev. Lett. \textbf{104}, 100503 (2010).

\end{thebibliography}
\end{document}